\documentclass[conference]{IEEEtran} 
\IEEEoverridecommandlockouts
\usepackage{cite}
\usepackage{url}
\usepackage{amsmath,amssymb,amsfonts}
\usepackage{algorithm}
\usepackage{algorithmic}
\usepackage{graphicx}
\usepackage{float}
\usepackage{subfig}
\graphicspath{{pics/}}
\newcommand{\figref}[1]{Fig.~\ref{#1}}
\usepackage{makecell}
\usepackage{threeparttable}
\usepackage{textcomp}
\usepackage{xcolor}
\usepackage{fancyhdr}
\def\BibTeX{{\rm B\kern-.05em{\sc i\kern-.025em b}\kern-.08em
    T\kern-.1667em\lower.7ex\hbox{E}\kern-.125emX}}

\title{Fluid Antenna Grouping Index Modulation\\ Design for MIMO Systems\\
\thanks{
This paper is supported in part by National Natural Science Foundation of China Program(62371291, 62271316, 62101322), National Key R\&D Project of China (2023YFF0904603), the Fundamental Research Funds for the Central Universities and Shanghai Key Laboratory of Digital Media Processing (STCSM 18DZ2270700). 

The corresponding author is Yin Xu (e-mail: xuyin@sjtu.edu.cn).}
}

\author{

    \IEEEauthorblockN{Xinghao Guo\IEEEauthorrefmark{1}, Yin Xu\IEEEauthorrefmark{1}, Dazhi He\IEEEauthorrefmark{1}, Cixiao Zhang\IEEEauthorrefmark{1},\\Wenjun Zhang\IEEEauthorrefmark{1}, \textit{Fellow, IEEE}, and Yi-yan Wu\IEEEauthorrefmark{2}, \textit{Life Fellow, IEEE}}
    \IEEEauthorblockA{
    \IEEEauthorrefmark{1} Cooperative Medianet Innovation Center (CMIC), Shanghai Jiao Tong University, Shanghai 200240, China \\
    \IEEEauthorrefmark{2} Wireless Technology Research Department, Communications Research Centre, Ottawa ON K2K 2Y6, Canada \\ 
    Email: \{guoxinghao, xuyin, hedazhi, cixiaozhang, zhangwenjun\}@sjtu.edu.cn, yiyan.wu@ieee.org
    }
    
}

\begin{document}

\fancyhf{}
\pagestyle{empty}
\thispagestyle{empty}

\maketitle

\begin{abstract}
     Index modulation (IM) significantly enhances the spectral efficiency of fluid antennas (FAs) enabled multiple-input multiple-output (MIMO) systems, which is named FA-IM. However, due to the dense distribution of ports on the FA, the wireless channel exhibits a high spatial correlation, leading to severe performance degradation in the existing FA-IM-assisted MIMO systems. To tackle this issue, this paper proposes a novel fluid antenna grouping index modulation (FA-GIM) scheme to mitigate the high correlation between the activated ports. Specifically, considering the characteristics of the FA two-dimensional (2D) surface structure and the spatially correlated channel model in FA-assisted MIMO systems, a block grouping method is adopted, where adjacent ports are assigned to the same group. Consequently, different groups independently perform port index selection and constellation symbol mapping, with only one port being activated within each group during each transmission interval. Then, a closed-form average bit error probability (ABEP) upper bound for the proposed scheme is derived. Numerical results show that, compared to state-of-the-art schemes, the proposed FA-GIM scheme consistently achieves significant bit error rate (BER) performance gains under various conditions. The proposed scheme is both efficient and robust, enhancing the performance of FA-assisted MIMO systems.
\end{abstract}

\begin{IEEEkeywords}
    Index modulation, fluid antenna, MIMO, spatial correlation, grouping, average bit error probability.
\end{IEEEkeywords}

\section{Introduction} \label{Sec-Intro}
\thispagestyle{empty}
\IEEEPARstart{W}{ith} the rapid increase of wireless devices, future wireless communication networks are anticipated to attain higher energy efficiency and spectrum efficiency. Numerous key technologies, including advanced coding and modulation \cite{NUC-0}, resource allocation \cite{ResAlloc-0}, and notably multiple-input multiple-output (MIMO) \cite{MIMO-0}, have been extensively applied in various communication networks to offer enormous data speeds and stable connectivity. MIMO technology, such as multi-user MIMO and massive MIMO, creates additional bandwidth by deploying multiple antennas to utilize spatial resources. The channels enabled by multiple antennas bring significant diversity and multiplexing advantages, fundamentally expanding communication capacity. Existing MIMO systems employ fixed position antennas (FPAs), where the spacing between antennas is typically not less than half a wavelength. 


An emerging and promising reconfigurable antenna technology known as fluid antenna systems (FAS) has recently attracted widespread attention \cite{FAS}. FAS refers to systems where antennas utilize flexible conductive materials, such as liquid metals or ionized solutions, and stepper motors \cite{LiquidAnt,MotorAnt}. Unlike conventional MIMO systems, FAS can be controlled via software to adjust the shape and position of FA, thereby reconfiguring the gain, radiation patterns, operating frequency, or other radiation characteristics \cite{MA}. \cite{FAS-PS} devised a port selection algorithm for FA in single-input single-output (SISO) systems by observing the channels of several ports. \cite{FAS-MIMO} studied the FAS-MIMO system from an information theory perspective and designed the corresponding joint optimization algorithm for port selection, beamforming, and power allocation.


As another promising technology, index modulation (IM) leverages the index of entities, such as antennas, subcarriers, time slots, etc., to convey additional information \cite{IM}. IM can reduce the number of radio frequency (RF) chains in communication systems, thus enhancing energy efficiency. IM has been integrated into conventional MIMO systems, termed spatial modulation (SM), where additional information bits are mapped to the index of the activated antenna \cite{SM}. However, SM systems have the following limitations: the number of transmit antennas must be a power of two, and the transmission rate grows logarithmically rather than linearly with the number of antennas. Generalized spatial modulation (GSM) scheme was proposed to relax the limitations above by activating multiple transmit antennas simultaneously and mapping information bits to the index of antenna combinations \cite{GSM}. Therefore, GSM offers more flexible antenna configurations and higher spectral efficiency. Nevertheless, the GSM scheme also introduces correlation among the transmit antennas, which will cause a degradation in system performance. \cite{G-GSM} explored grouping methods to address this issue in GSM systems.

FAS and IM share similar physical characteristics, specifically in selecting transmission entities. A natural integration scheme, fluid antenna index modulation (FA-IM), was proposed in \cite{IM-FA}, where a FA port is activated during each transmission interval to convey additional information. In \cite{FA-IM-NN}, neural network is employed to achieve fast classification of index patterns in FA-IM systems. Based on the traditional GSM scheme, \cite{FA-IM} applied the FA-IM scheme to MIMO systems, where multiple ports are simultaneously activated to achieve higher multiplexing gains. Compared to the FAS-MIMO scheme in \cite{FAS-MIMO}, the FA-IM scheme significantly enhances the spectral efficiency of the MIMO system. The FA-IM scheme also obtains a higher bit error rate (BER) performance gain when compared under the same transmission rate. However, \cite{FA-IM} neglects the issue of correlated channel between multiple activated ports. The ports in FAS are densely distributed, with adjacent ports exhibiting a strong spatial correlation, which has a non-negligible impact. 

To fill this gap, this paper proposes an innovative FA grouping IM (FA-GIM) scheme. The ports on the FA are evenly divided into multiple groups, with each group independently implementing index modulation and data symbol modulation. Based on the spatial correlation and distribution structure characteristics of the ports, a block grouping method is employed, where adjacent ports with high correlation are assigned to the same group. The corresponding closed-form average bit error probability (ABEP) is derived. Simulation results are provided to highlight the effectiveness and robustness of the proposed FA-GIM scheme. 


\begin{figure*}[htbp]
    \centerline{\includegraphics[width=0.7\textwidth]{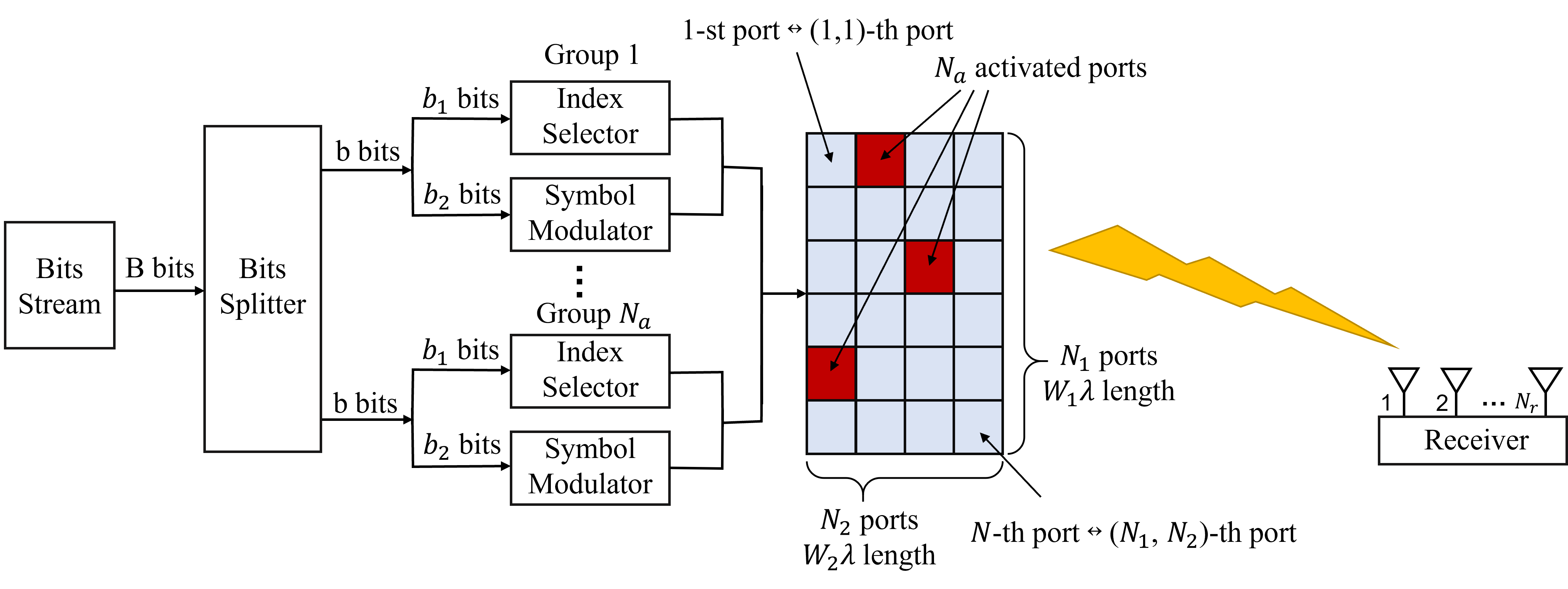}}
    
    \captionsetup{justification=raggedright, singlelinecheck=false}
    
    \caption{Block diagram of the proposed FA-GIM system.}
    \label{fig:FA-GIM system}
\end{figure*}

\section{System Model} \label{Sec-Model}
The proposed FA-GIM system is depicted in \figref{fig:FA-GIM system}, where the transmitter is equipped with a fluid antenna and the receiver is equipped with $N_{r}$ FPAs. Assume the FA with $N$ ports occupies a two-dimensional (2D) surface with a range of $W= \lambda W_{1} \times \lambda W_{2}$, where $\lambda$ is the wavelength of the carrier, and $\lambda W_{1}$ and $\lambda W_{2}$ denote the length of the surface along the vertical and horizontal directions, respectively. A grid structure is considered, with $N_{1}$ and $N_{2}$ ports uniformly distributed along the vertical and horizontal directions, respectively. So we have $N=N_{1} \times N_{2}$. During each transmission interval, the transmitter activates $N_{a}$ ports out of $N$ ports.

\subsection{Transmission}
A total of $B$ information bits enter the transmitter during each transmission interval. The transmitter selects and activates $N_{a}$ ports out of $N$ ports based on these $B$ bits to carry $N_{a}$ data symbols. Specifically, these $B$ bits are then equally divided into $N_{a}$ groups, with each group containing $b$ bits, i.e., $B=b \times N_{a}$. Correspondingly, the $N$ ports of the FA are divided into $N_{a}$ groups, with each group containing $N_{p}$ ports, i.e., $N=N_{p} \times N_{a}$. 

The $b$ bits input to each group are further divided into two parts. In the first part, a total of $b_{1}$ bits are used to select and activate one port out of $N_{p}$ ports, meaning that $b_{1}$ bits are mapped to the index of the activated port. Assume that $N_{p}$ is a power of 2, so $b_{1}=\log_2 N_{p} = \log_2 \frac{N}{N_{a}}$. Let $I_{g}$ represents the port index mapped from the $g$-th group, i.e., $I_{g} \in \{1,2,...,N\}$ for $g \in \{1,2,...,N_{a}\}$. Then, the set of port indices selected by the total $B$ bits can be represented as $\mathcal{I}= \{I_{1},...,I_{g},...,I_{N_{a}}\}$. The remaining $b_{2}$ bits are modulated to symbol $s_{g}$ using a $M$-ary constellation, with $b_{2}=\log_2 M$. Thus, the transmitted signal vector $\mathbf{x} \in \mathbb{C}^{N \times 1}$ can be represented as:
\begin{equation}
\label{eq-signal-t}
\mathbf{x}=\sum_{g=1}^{N_{a}} s_{g} \mathbf{e}_{I_{g}},
\end{equation}%
where $\mathbf{e}_{I_{g}} $ denotes an $N$-dimensional standard basis vector with the $I_{g}$-th position being 1 and all other positions being 0. The total number of transmitted bits $B$ in the FA-GIM scheme can be expressed as
\begin{equation}
\label{bpcu-FA-GIM}
B= N_{a} \times (\log_2 \frac{N}{N_{a}} + \log_2 M).
\end{equation}%

In contrast, the FA-IM scheme proposed in \cite{FA-IM} does not include the grouping operation. Specifically, the $B^{\prime}$ information bits entering the transmitter are divided into two parts. The first part is used to select and activate $N_{a}$ ports out of $N$ ports, which means it is mapped to the index of the port combination. Therefore, during each transmission interval, the number of bits in the first part is $\lfloor \log_2 \binom{N}{N_{a}} \rfloor$, where $\binom{\cdot}{\cdot}$ and $\lfloor \cdot \rfloor$ denote the binomial coefficient and the floor operation, respectively. The bits in the second part are modulated to $N_{a}$ data symbols using a $M$-ary constellation. Therefore, during each transmission interval, the number of bits in the second part is $N_{a} \times \log_2 M$. So, in the FA-IM scheme, the total number of information bits $B^{\prime}$ conveyed per transmission interval can be expressed as
\begin{equation}
\label{bpcu-FA-IM}
B^{\prime}= \left \lfloor \log_2 \binom{N}{N_{a}} \right \rfloor + N_{a} \times \log_2 M.
\end{equation}%

\subsection{Spatial Correlation Model}  \label{Sec-Model-SS-Corr}
The spatial position of the $n$-th port located on the 2D surface of the FA is denoted as $\mathbf{t}_{n} = [x_{n}, y_{n}]$ for $n \in \{1,2,...,N\}$. Considering a three-dimensional (3D) environment under rich scattering, the spatial correlation between the $n$-th port and the $m$-th port has been provided in \cite{FAS-MIMO}:
\begin{equation}
\label{eq-corr-coeff}
J_{n,m}= \mathrm{sinc}(k \|\mathbf{t}_n-\mathbf{t}_m\|_2),
\end{equation}%
where $\mathrm{sinc}(z)=\frac{ \mathrm{sin}(z) }{z}$, $k=\frac{2 \pi}{\lambda}$, and $\| \cdot \|_2$ is the $\ell_2$ norm. As  seen in \eqref{eq-corr-coeff}, it is essential to establish a functional mapping between the positions of ports on the 2D surface and their labels, which will be elaborated upon in Section \ref{Sec-Prop}. With the spatial correlation between ports, the transmitter correlation matrix $\mathbf{J}_{t}$ can be expressed as
\begin{equation}
\label{eq-corr-mat}
\mathbf{J}_{t}=\left[
\begin{array}{cccc}
J_{1,1}&J_{1,2}&\ldots&J_{1,N}\\
J_{2,1}&J_{2,2}&\ldots&J_{2,N}\\
\vdots&\vdots&\ddots&\vdots\\
J_{N,1}&J_{N,2}&\ldots&J_{N,N}
\end{array}
\right].
\end{equation}%
Since $J_{n,m}= J_{m,n}$, \eqref{eq-corr-mat} can be further decomposed into $\mathbf{J}_{t}= \mathbf{U}_{t} \mathbf{\Lambda}_{t} \mathbf{U}_{t}^{H}$, where $\mathbf{U}_{t}$ is an $N \times N$ matrix whose columns are the eigenvector of $\mathbf{J}_{t}$ and $\mathbf{\Lambda}_{t}= \mathrm{diag}\left(\lambda_{1}^{t},\ldots,\lambda_{N}^{t}\right)$ is an $N \times N$ diagonal matrix whose diagonal entries are the corresponding eigenvalues.

Assuming no path loss, the equivalent channel matrix from the $N$ ports of the transmit FA to $N_{r}$ receive FPAs can be represented as
\begin{equation}
\label{corr-chan}
\mathbf{H}= \mathbf{J}_{r}^{\frac{1}{2}} \mathbf{G} \mathbf{J}_{t}^{\frac{1}{2}} = \mathbf{J}_{r}^{\frac{1}{2}} \mathbf{G} \sqrt{\mathbf{\Lambda}_{t}^{H}} \mathbf{U}_{t}^{H},
\end{equation}%
where $\mathbf{J}_{r} \in \mathbb{C}^{N_{r} \times N_{r}}$ is the receiver correlation matrix and $\mathbf{G} \in \mathbb{C}^{N_{r} \times N}$ is the uncorrelated channel matrix, whose elements are i.i.d random values following $\mathcal{CN} (0,1)$. Note that $\mathbf{J}_{r}$ and $\mathbf{J}_{t}$ are independently calculated, and IM is implemented only at the FA of the transmitter. The performance of the FA-IM scheme and the FA-GIM scheme is primarily influenced by the transmitter-side correlation. Therefore, this paper sets $\mathbf{J}_{r} = \mathbf{I}_{N_{r}}$ with $\mathbf{I}_{N_{r}}$ being the identical matrix of size $N_{r}$ to ignore the correlation at the receiver side, which is consistent with the condition in \cite{G-GSM}. The correlated channel matrix can be simplified as
\begin{equation}
\label{channel}
\mathbf{H}= \mathbf{G} \sqrt{\mathbf{\Lambda}_{t}^{H}} \mathbf{U}_{t}^{H}.
\end{equation}%

\subsection{Detection}
With the correlated channel matrix, the received signal $\mathbf{y} \in \mathbb{C}^{N_{r} \times 1}$ can be written as
\begin{equation}
\label{y_receive}
\mathbf{y}= \mathbf{H} \mathbf{x} + \mathbf{w},
\end{equation}%
where $\mathbf{w} \in \mathbb{C}^{N_{r} \times 1} \sim \mathcal{CN}(0, N_{0} \mathbf{I}_{N_{r}})$ is the additive white Gaussian noise.

Assuming that the perfect channel state information (CSI) is available at receiver, the optimal maximum likelihood (ML) detector can be expressed as
\begin{equation}
\begin{aligned}
\label{eq-detect-ML}
    (\hat{\mathcal{I}},\hat{\mathbf{s}})&= \arg \underset{\mathcal{I},\mathbf{s}}{\min} \left \| \mathbf{y} - \mathbf{H} \mathbf{x} \right \|^2\\
    &=\arg \underset{\mathcal{I},\mathbf{s}}{\min} \left \| \mathbf{y} - \mathbf{H} \sum_{g=1}^{N_{a}} s_{g} \mathbf{e}_{I_{g}} \right \|^2,
\end{aligned}
\end{equation}%
where $\mathbf{s}=[s_{1},...,s_{g},...,s_{N_a}]^{T}$, which comprises $N_a$ modulated data symbols. 

\section{Grouping Method and Mapping From Spatial Positions to Labels}  \label{Sec-Prop}

\begin{figure}[htbp]
    \centerline{\includegraphics[width=0.49\textwidth]{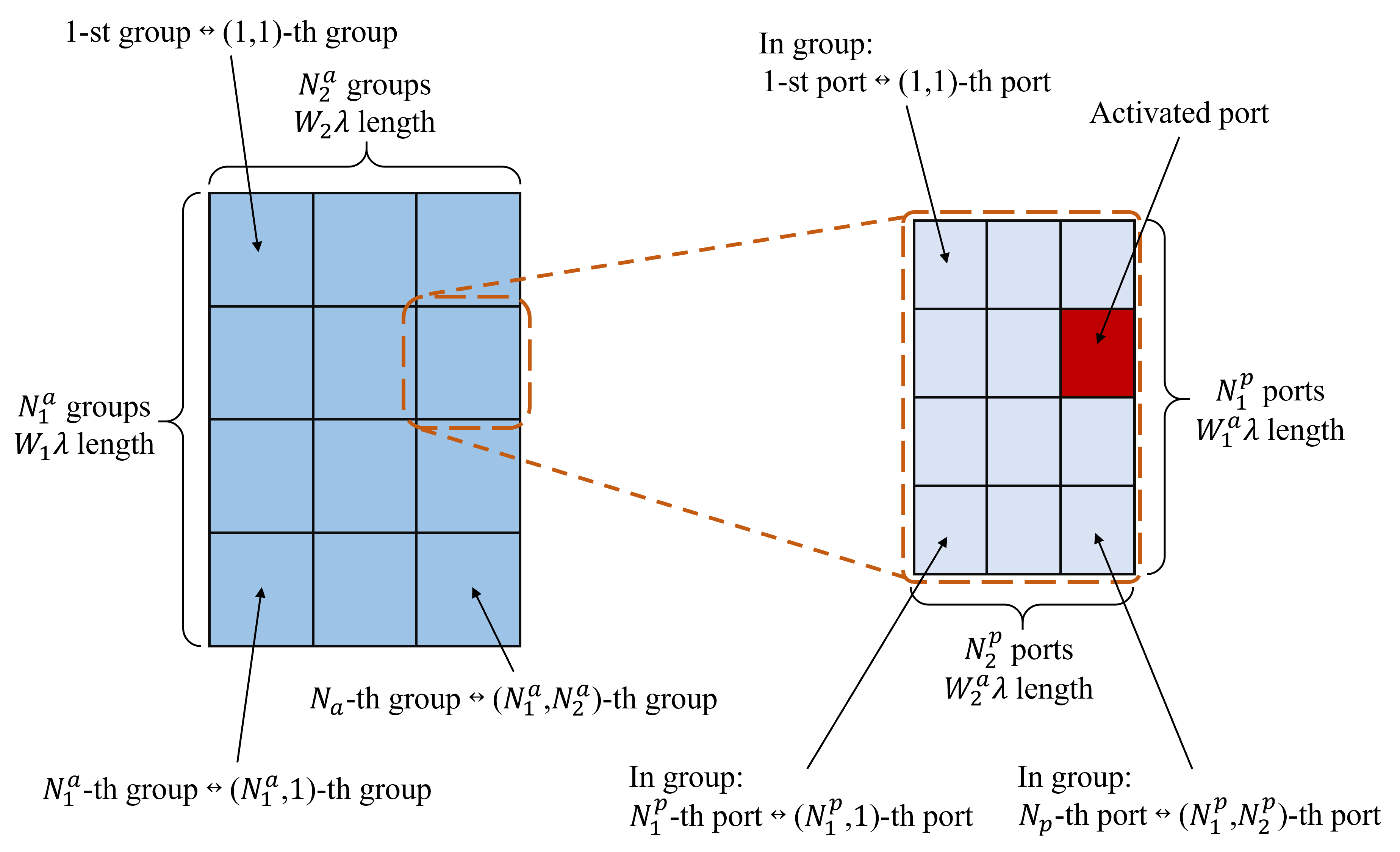}}
    \caption{Illustration of the block grouping method and mapping from positions to labels.}
    \label{fig:grouping}
\end{figure}

According to the spatial correlation calculation formula \eqref{eq-corr-coeff} between ports distributed on the 2D surface, it can be observed that the smaller the geometric distance between ports, the higher the spatial correlation. Considering the operational mechanism of FA-IM, it is reasonable to disperse the activated ports as much as possible to mitigate spatial correlation. Therefore, this paper adopts a block grouping method, as shown in \figref{fig:grouping}, to partition $N$ ports into $N_{a}$ groups. One port is selected and activated based on the input $b_{1}$ bits within each group.

Specifically, the whole transmit FA surface is uniformly divided into $N_{a}= N^{a}_{1} \times N^{a}_{2}$ groups, where $N^{a}_{1}$ and $N^{a}_{2}$ denote the number of groups along vertical and horizontal directions, respectively. Let $\lambda W^{a}_{1}= \lambda W_{1}/ N^{a}_{1}$ and $\lambda W^{a}_{2}= \lambda W_{2}/ N^{a}_{2}$ denote the length of a group along vertical and horizontal directions, respectively. The $(a_{1}, a_{2})$-th group refers to the group located in the $a_{1}$-th row and $a_{2}$-th column for $a_{1} \in \{1,2,...,N^{a}_{1}\}$ and $a_{2} \in \{1,2,...,N^{a}_{2}\}$. The $N_{a}$ groups are sequentially labeled in the order from top to bottom and left to right, as shown in \figref{fig:grouping}. Thus, the mapping function from the location to the label of the $g$-th group can be expressed as 
\begin{equation}
\label{label-group}
g(a_{1},a_{2})= (a_{2}-1)N^{a}_{1} + a_{1}, \quad g \in \{1,2,...,N_{a}\}.
\end{equation}%

Based on the block grouping method, the number of ports $N_{p}$ within each group satisfies $N_{p}= N^{p}_{1} \times N^{p}_{2}$, where $N^{p}_{1}$ and $N^{p}_{2}$ denote the number of ports along vertical and horizontal directions, respectively. Let $\lambda W^{p}_{1}= \lambda W^{a}_{1}/ N^{p}_{1} = \lambda W_{1}/ N_{1}$ and $\lambda W^{p}_{2}= \lambda W^{a}_{2}/ N^{p}_{2} = \lambda W_{2}/ N_{2}$ denote the length occupied by a port along the vertical and horizontal directions, respectively. The $(p_{1}, p_{2})$-th port refers to the port located in the $p_{1}$-th row and $p_{2}$-th column  within a given group for $p_{1} \in \{1,2,...,N^{p}_{1}\}$ and $p_{2} \in \{1,2,...,N^{p}_{2}\}$. The $N_{p}$ ports are also sequentially labeled in the order from top to bottom and left to right, as shown in \figref{fig:grouping}. Thus, the mapping function from the location to the label of the $k$-th port within a given group can be expressed as 
\begin{equation}
\label{label-port-group}
k(p_{1},p_{2})= (p_{2}-1)N^{p}_{1} + p_{1}, \quad k \in \{1,2,...,N_{p}\}.
\end{equation}%

Given the mapping functions \eqref{label-group} and \eqref{label-port-group} above, from the perspective of the entire 2D surface, the label $n$ of the $k$-th port within the $g$-th group can be expressed as 
\begin{equation}
\label{label-port-FA}
n(g,k)= (g-1)N_{p} + k, \quad n \in \{1,2,...,N\}.
\end{equation}%

With the mapping functions from locations to labels, it becomes easy to represent the spatial positions of ports on the 2D surface. The port located at the top-left corner of the 2D surface is chosen as the reference position $\mathbf{t}_{ref}$, which is denoted as $\mathbf{t}_{ref} = \mathbf{t}_{1} = [0,0]$. Similarly, for the $g$-th group, the port located at the top-left corner within the group is chosen as the reference position $\mathbf{t}^{g}_{ref}$, which is denoted as $\mathbf{t}^{g}_{ref} = \mathbf{t}^{g}_{1} = [(a_{1}-1) \lambda W^{a}_{1},(a_{2}-1) \lambda W^{a}_{2}]$. Thus, the spatial position $\mathbf{t}_{n}$ of the $n$-th port on the 2D surface, which corresponds to the $k$-th port within the $g$-th group, can be represented as 
\begin{equation}
\label{position-port}
\mathbf{t}_{n}= \mathbf{t}^{g}_{k}= \mathbf{t}^{g}_{ref} + [(p_{1}-1) \lambda W^{p}_{1},(p_{2}-1) \lambda W^{p}_{2}].
\end{equation}%

Additionally, the block grouping method designed for FA 2D surface structure and the corresponding mapping function from spatial positions to labels of ports can be easily transformed to obtain counterparts for one-dimensional (1D) FA structures, i.e., by simply setting $N_{2}=1$ and $N^{a}_{2}=1$.

\begin{table}[htbp]
\renewcommand\arraystretch{1.2}
\centering
\caption{Transmitted signal vector $\mathbf{x}$ of the proposed FA-GIM scheme and the FA-IM scheme for $N=4=4 \times 1, N_{a}=2=2 \times 1$.}
\label{tab:grouping}
\begin{tabular}{c c c}
\hline \hline
\textbf{Port index bits}  & \textbf{FA-IM in \cite{FA-IM}}  & \textbf{FA-GIM with block grouping} \\ 
\hline
$[0 \thickspace 0]$  & $[s_{1},s_{2},0,0]^{T}$  & $[s_{1},0,s_{2},0]^{T}$  \\ 
$[0 \thickspace 1]$  & $[0,s_{1},s_{2},0]^{T}$  & $[s_{1},0,0,s_{2}]^{T}$  \\ 
$[1 \thickspace 0]$  & $[0,0,s_{1},s_{2}]^{T}$  & $[0,s_{1},s_{2},0]^{T}$  \\ 
$[1 \thickspace 1]$  & $[s_{1},0,s_{2},0]^{T}$  & $[0,s_{1},0,s_{2}]^{T}$  \\ 
\hline \hline
\end{tabular}
\end{table}

For clarity, considering a simple example of $N=4=N_{1} \times N_{2}=4 \times 1$ and $ N_{a}=2=N^{a}_{1} \times N^{a}_{2}=2 \times 1$, Table \ref{tab:grouping} lists a possible ensemble of transmitted signal vector $\mathbf{x}$ in \eqref{eq-signal-t} to distinctly highlight the difference between the proposed FA-GIM scheme, which incorporates the block grouping method, and the FA-IM scheme in \cite{FA-IM}.

\section{Performance Analysis}  \label{Sec-Perf}

This section derives a closed-form average bit error probability (ABEP) upper bound for the proposed FA-GIM system. Based on the ML detector in \eqref{eq-detect-ML}, the conditional pairwise error probability (CPEP) is given as
\begin{equation}
\begin{aligned}
    \label{eq-CPEP-ori}
    P\{ (\mathcal{I} \to \hat{\mathcal{I}}, \mathbf{s} \to \hat{\mathbf{s}}) | \mathbf{H} \}
    &=P\{ (\mathbf{x} \to \hat{\mathbf{x}}) | \mathbf{H} \}\\
    &=P\left \{ \left \| \mathbf{y} - \mathbf{H} \mathbf{x} \right \|^2 \geq \left \| \mathbf{y} - \mathbf{H} \hat{\mathbf{x}} \right \|^2\right \}\\
    &=Q\left (\sqrt{\frac{\left \| \mathbf{H} \Psi \right \|^2}{2N_0}}\right )=Q\left(\sqrt{\frac{\Gamma}{2N_0}}\right),
\end{aligned}
\end{equation}%
where $\Psi = \mathbf{x} - \hat{\mathbf{x}}$, $\Gamma = \| \mathbf{H} \Psi \|^2$, and $Q(\cdot)$ is the Gaussian $Q$-function. Based on the approximate upper bound of the $Q$-function \cite{ref-Qfun}, the CPEP can be approximated as
\begin{equation}
\begin{aligned}
    \label{eq-CPEP-appr}
    & P\{ (\mathcal{I} \to \hat{\mathcal{I}}, \mathbf{s} \to \hat{\mathbf{s}}) | \mathbf{H} \}\\
    &\approx \frac{1}{6}\exp \left (-\frac{\Gamma}{N_0} \right )+\frac{1}{12}\exp \left (-\frac{\Gamma}{2N_0} \right )+\frac{1}{4}\exp \left (-\frac{\Gamma}{4N_0} \right ).
\end{aligned}
\end{equation}%
Furthermore, the unconditional pairwise error probability (UPEP) can be expressed as
\begin{equation}
\begin{aligned}
    \label{eq-UPEP-appr}
    & P\{ (\mathcal{I} \to \hat{\mathcal{I}}, \mathbf{s} \to \hat{\mathbf{s}}) \}\\
    &\approx \frac{1}{6}M_{\Gamma} \left (-\frac{1}{N_0} \right )+\frac{1}{12}M_{\Gamma} \left (-\frac{1}{2N_0} \right )+\frac{1}{4}M_{\Gamma} \left (-\frac{1}{4N_0} \right ).
\end{aligned}
\end{equation}%
where $M_\Gamma(x)=\mathbb{E}_{\Gamma}[\exp (x\Gamma)]$ is the moment generating function (MGF) of $\Gamma$.

Since the format of the transmitted signal vector $\mathbf{x}$ in \eqref{eq-signal-t} resembles that in conventional GSM, some of the analytical results in \cite{ref-ABEP} can be integrated with some modifications to obtain 
\begin{equation}
\begin{aligned}
    \label{eq-MGF-ori}
    M_\Gamma(x)= \frac12 \frac{ \exp \left (x \mathbf{u}_{\mathbf{H}}^{H} \Lambda \left (\mathbf{I} - x \mathbf{L}_{\mathbf{H}} \Lambda \right )^{-1} \mathbf{u}_{\mathbf{H}} \right ) }{ \left | \mathbf{I} - x \mathbf{L}_{\mathbf{H}} \Lambda \right | },
\end{aligned}
\end{equation}%
where $| \cdot |$ denotes the determinant of a matrix, $\mathbf{I}$ is the identity matrix, $\mathbf{u}_{\mathbf{H}} = u_{\mathbf{H}} (\mathbf{I}_{N_{r}} \otimes \mathbf{J}_t)^{\frac12} \text{vec}(\mathbf{1}_{N_{r}N_{t}})$, $\Lambda = \mathbf{I}_{N_{r}} \otimes \Psi \Psi^{H}$, and $\mathbf{L}_{\mathbf{H}} = \sigma_{\mathbf{H}}^{2} \mathbf{J}_{r} \otimes \mathbf{J}_{t}$, where $\otimes$ is the Kronecker product, $\text{vec}(\cdot)$ is the vectorisation operator, and $\mathbf{1}_{N_{r}N_{t}}$ is an $N_{r} \times N_{t}$ all one matrix. Under the channel assumptions in Section \ref{Sec-Model-SS-Corr}, it follows that $u_{\mathbf{H}} = 0$, $\sigma_{\mathbf{H}}^{2} = 1$, and $\mathbf{L}_{\mathbf{H}} = \mathbf{I}_{N_{r}} \otimes \mathbf{J}_{t}$, hence \eqref{eq-MGF-ori} can be simplified as
\begin{equation}
    \label{eq-MGF-simp}
    M_\Gamma(x)= \frac12 \frac{1}{ \left | \mathbf{I} - x \mathbf{J}_{t} \Psi \Psi^{H} \right |^{N_{r}} }.
\end{equation}%

The ABEP upper bound of the proposed FA-GIM system can be expressed as
\begin{equation}
    \label{eq-ABEP-ori}
    \text{ABEP} \leq \frac{1}{2^{B}}\sum_{\mathcal{I}, \mathbf{s}} \sum_{\hat{\mathcal{I}}, \hat{\mathbf{s}}} \frac{ P\{ (\mathcal{I} \to \hat{\mathcal{I}}, \mathbf{s} \to \hat{\mathbf{s}}) \} e(\mathcal{I} \to \hat{\mathcal{I}}, \mathbf{s} \to \hat{\mathbf{s}}) }{B},
\end{equation}%
\begin{figure*}[htbp] 
\centering
\begin{equation}	    
\text{ABEP}\leq \frac{1}{2^{B} B} \sum_{\mathcal{I}, \mathbf{s}} \sum_{\hat{\mathcal{I}}, \hat{\mathbf{s}}} e(\mathcal{I} \to \hat{\mathcal{I}}, \mathbf{s} \to \hat{\mathbf{s}}) \left ( \frac{1}{12} \frac{1}{ \left | \mathbf{I} + \frac{1}{N_0} \mathbf{J}_{t} \Psi \Psi^{H} \right |^{N_{r}} } + \frac{1}{24} \frac{1}{ \left | \mathbf{I} + \frac{1}{2N_0} \mathbf{J}_{t} \Psi \Psi^{H} \right |^{N_{r}} } + \frac{1}{8} \frac{1}{ \left | \mathbf{I} + \frac{1}{4N_0} \mathbf{J}_{t} \Psi \Psi^{H} \right |^{N_{r}} } \right )
\label{eq-ABEP-final}
\end{equation}
\hrulefill
\end{figure*}
where $e(c,\mathbf{s}\to\hat{c},\hat{\mathbf{s}})$ represents the total number of erroneous bits for the corresponding pairwise error event. 

With the simplified MGF derived from \eqref{eq-MGF-simp}, substituting \eqref{eq-UPEP-appr} into \eqref{eq-ABEP-ori} yields \eqref{eq-ABEP-final}. Note that \eqref{eq-ABEP-final} does not require numerical evaluation of integrals. Section \ref{Sec-Simu} will verify that the derived ABEP upper bound in \eqref{eq-ABEP-final} is a tight bound for the proposed FA-GIM system.

\section{Simulation Results}  \label{Sec-Simu}

This section evaluates the proposed FA-GIM scheme BER performance and makes some comparisons. For the fairness of comparison, in the simulation, the total transmission power of the transmitter is set to 1, i.e., $\mathbb{E} [ \| \mathbf{x} \|^2 ] = 1$, and all schemes employ the optimal ML detector. 

\begin{figure}[htbp]
    \centering
	\subfloat[$N_{r}=2$\label{fig:BER11}]{
	\includegraphics[width=.45\columnwidth]{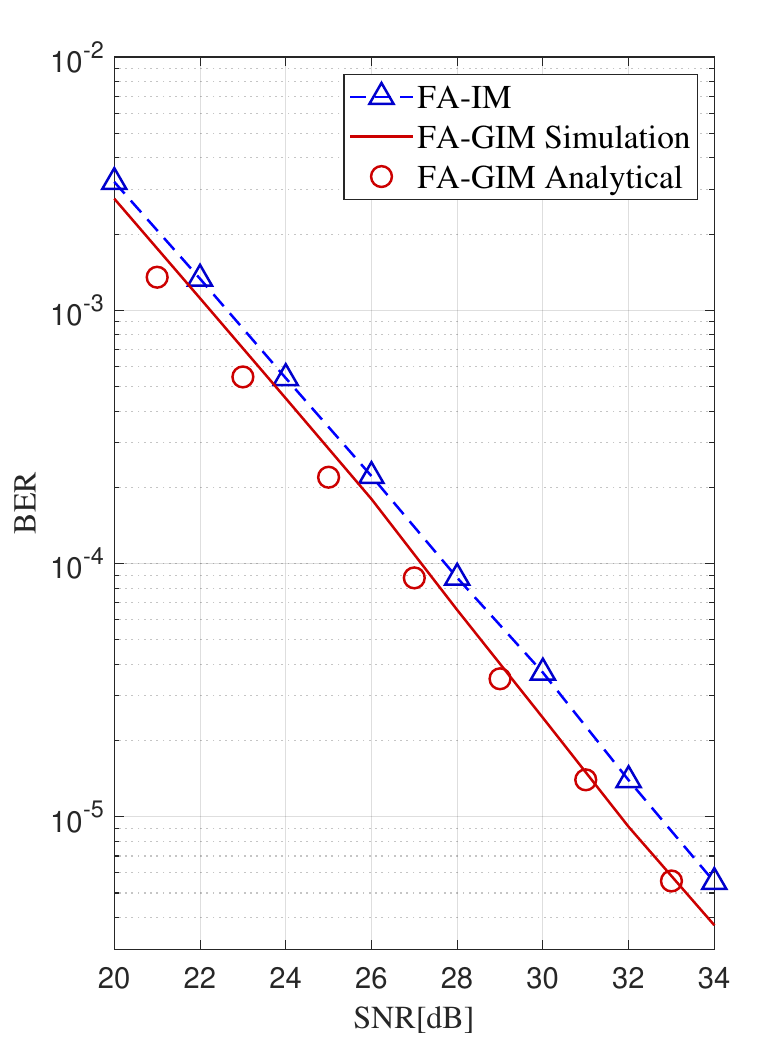}
    }
	\subfloat[$N_{r}=4$\label{fig:BER12}]{
	\includegraphics[width=.45\columnwidth]{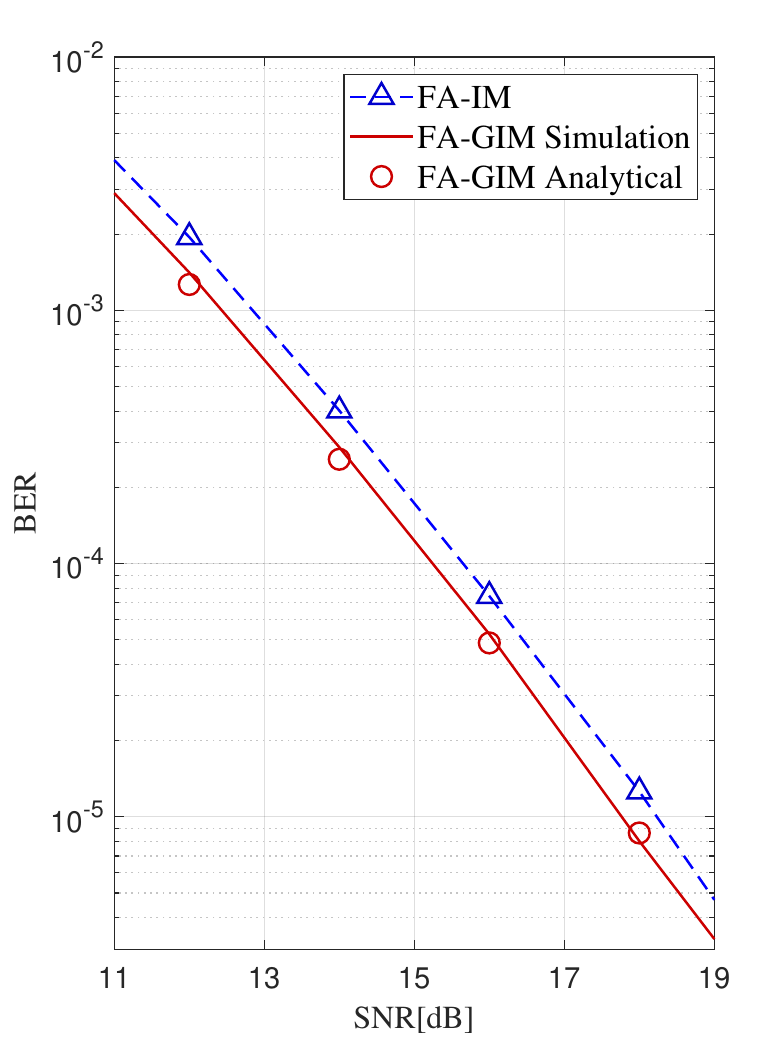}
	}\\
	\subfloat[$N_{r}=8$\label{fig:BER13}]{
	\includegraphics[width=.45\columnwidth]{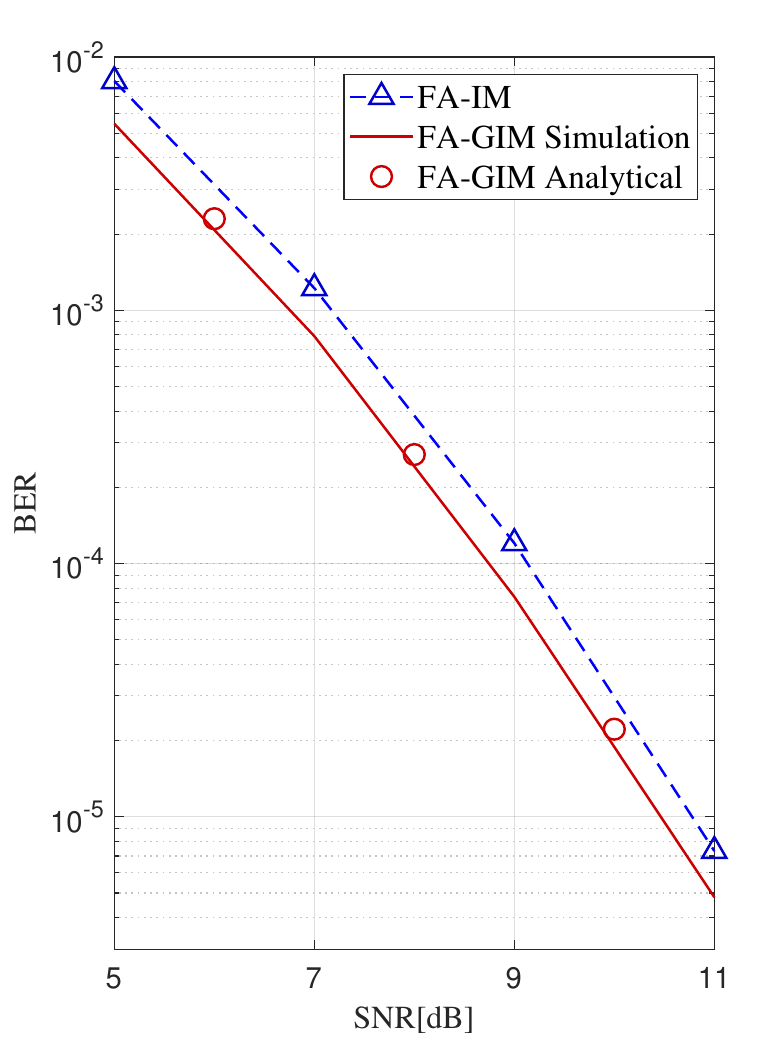}
	}
    \subfloat[$N_{r}=16$\label{fig:BER14}]{
	\includegraphics[width=.45\columnwidth]{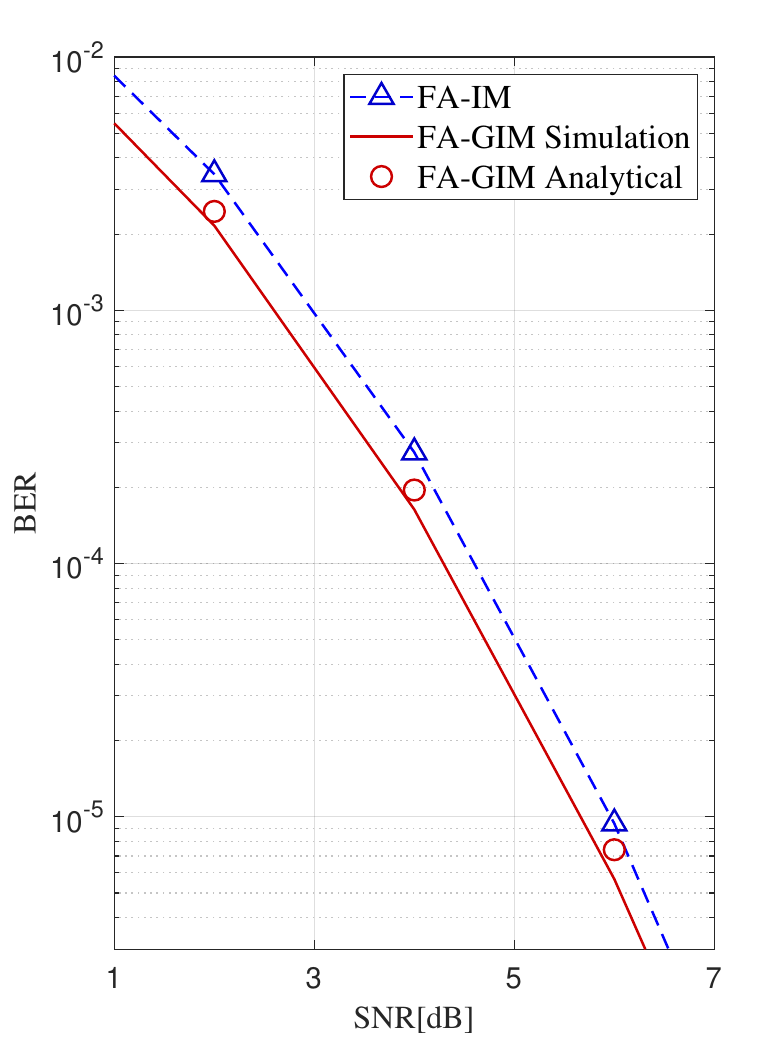}
	}
	\caption{BER performance comparisons between the proposed FA-GIM and FA-IM schemes with BPSK, $W_{1}=2, W_{2}=4, N=8=2 \times 4, N_{a}=2=1 \times 2$ and different $N_{r}$, i.e., $N_{r}=$ (a) $2$, (b) $4$, (c) $8$, (d) $16$.}
	\label{fig:BER_Nr}
\end{figure}

Firstly, \figref{fig:BER_Nr} illustrates the comparison of BER performance between the proposed FA-GIM scheme and the FA-IM scheme in \cite{FA-IM} at the same transmission rate. The number of receive antennas $N_{r}$ is varied to observe its impact on both schemes. We set the system configuration of $W_{1}=2, W_{2}=4, N=8=N_{1} \times N_{2}=2 \times 4, N_{a}=2=N^{a}_{1} \times N^{a}_{2}=1 \times 2$. Both schemes employ BPSK modulation, thus their transmission rates are equal with $B= B^{\prime}= 6$. It can be observed that as the number of receive antennas $N_{r}$ increases, the BER performance of both schemes improves. This trend is reasonable because increasing the number of receive antennas increases receive diversity gain. However, compared to the FA-IM scheme, the proposed FA-GIM scheme consistently achieves BER performance gains, regardless of variations in the number of receive antennas. Specifically, at a BER value of $10^{-5}$, the proposed FA-GIM scheme can provide approximately 0.9 dB, 0.5 dB, 0.35 dB, and 0.3 dB gains for $N_{r}=$ 2, 4, 8, and 16, respectively. Besides, as shown in \figref{fig:BER_Nr}, the theoretical ABEP curves of the proposed FA-GIM scheme match well with the Monte Carlo simulation results in the high signal-to-noise-ratio (SNR) region. The high spatial correlation among the ports on the transmitter FA makes distinguishing ports harder for the receiver. In this context, the proposed FA-GIM scheme can mitigate the correlation among the activated ports, leading to higher performance gains in scenarios with fewer receive antennas.

\begin{figure}[htbp]
    \centering
	\subfloat[$W_{1}=W_{2}=2.4$\label{fig:BER21}]{
	\includegraphics[width=.45\columnwidth]{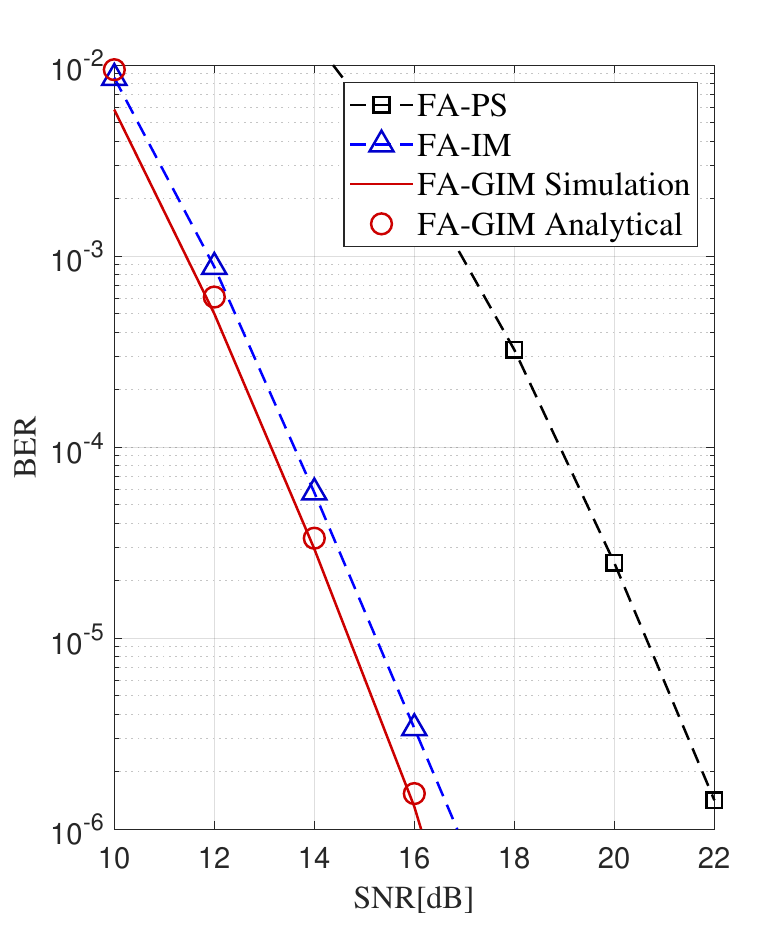}
    }
	\subfloat[$W_{1}=W_{2}=4.8$\label{fig:BER22}]{
	\includegraphics[width=.45\columnwidth]{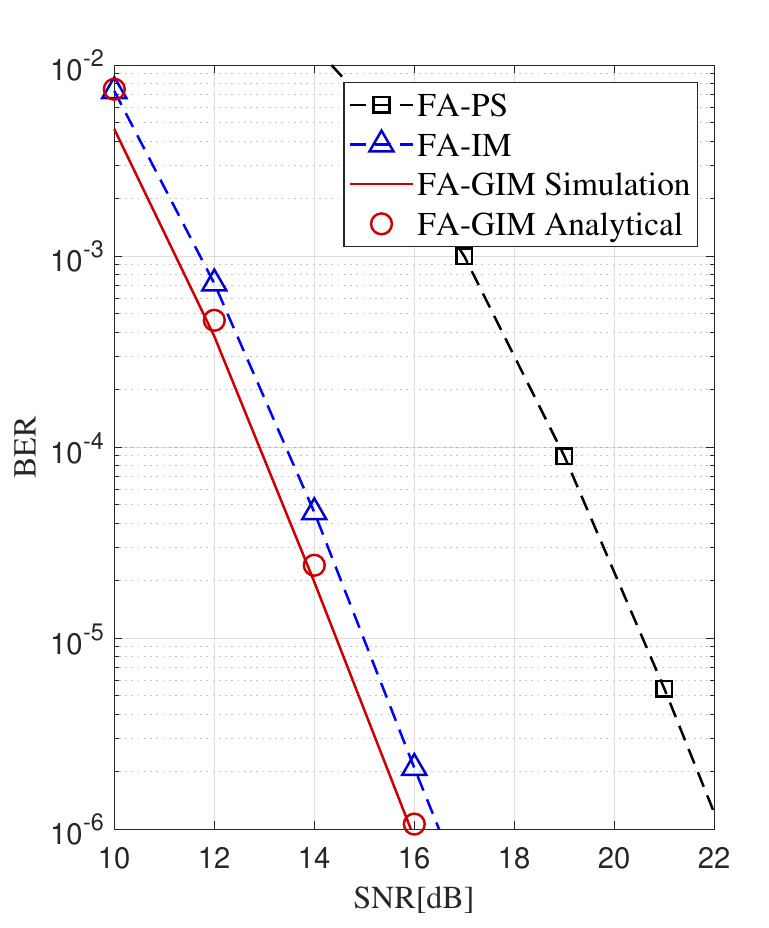}
	}\\
	\subfloat[$W_{1}=W_{2}=9.6$\label{fig:BER23}]{
	\includegraphics[width=.45\columnwidth]{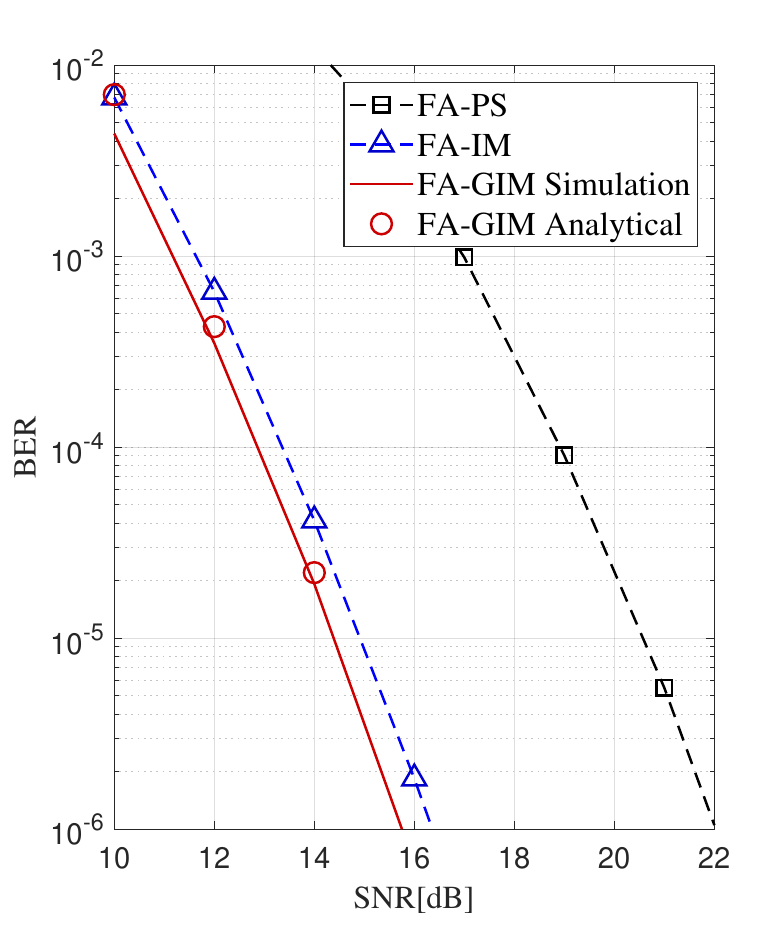}
	}
    \subfloat[$W_{1}=W_{2}=19.2$\label{fig:BER23}]{
	\includegraphics[width=.45\columnwidth]{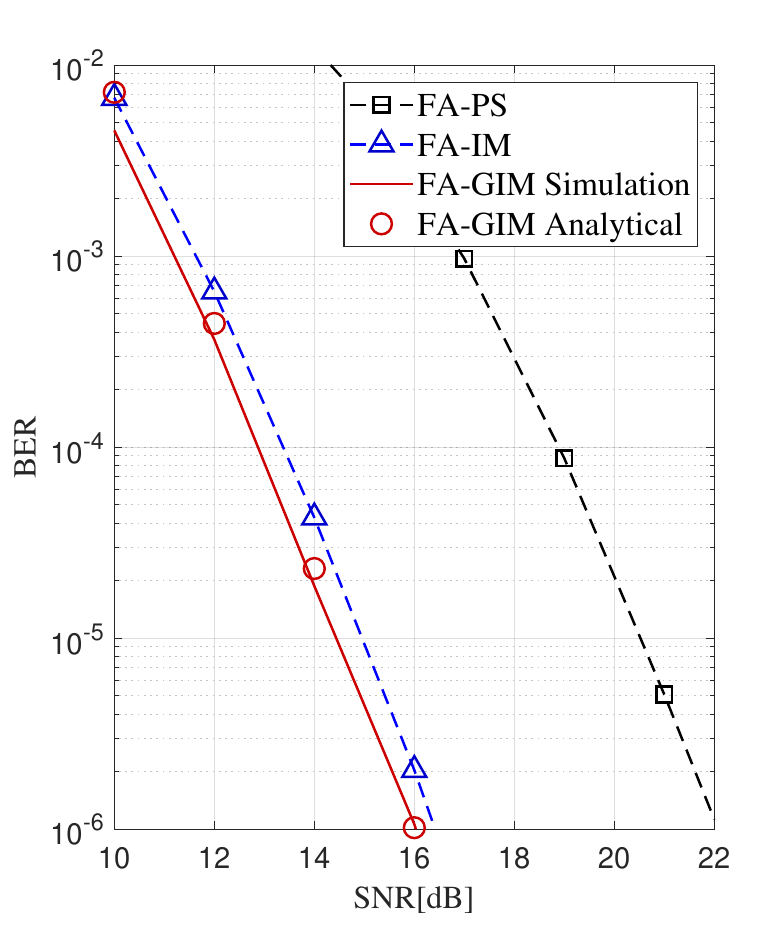}
	}
	\caption{BER comparison results for FA-GIM, FA-IM, and FA-PS schemes using 4-QAM, BPSK, and 16-QAM, respectively, with $N_R=8$, $N_{a}=4=2 \times 2$ and different port densities, i.e., $W_{1}=W_{2}=$ (a) $2.4$, (b) $4.8$, (c) $9.6$, (d) $19.2$ with $N=16=4 \times 4$ fixed.}
	\label{fig:BER_W}
\end{figure}

Next, \figref{fig:BER_W} investigates the BER performance comparison of schemes, including the existing FA port selection (FA-PS) scheme \cite{FAS-PS,FAS-MIMO}, under different port densities. With the number of ports $N=16=N_{1} \times N_{2}=4 \times 4$ fixed, the port density can be changed by varying the relative side length $W_{1}$ and $W_{2}$ of the FA 2D surface. To ensure fairness in the comparisons and to achieve consistent transmission rates as much as possible, all schemes use the system configuration $N_R=8$ and $N_{a}=4=N^{a}_{1} \times N^{a}_{2}=2 \times 2$. The FA-GIM scheme, FA-IM scheme, and FA-PS scheme employ 4-QAM, BPSK, and 16-QAM, respectively, resulting in transmission rates of $B= 16, B^{\prime}= 14$, and $B_{FA-PS}= 16$. As seen from \figref{fig:BER_W}, with increasing SNR, the analytical results gradually converge towards the simulation results of the proposed FA-GIM scheme. Besides, it can be observed that the proposed FA-GIM scheme consistently provides significant performance gains across different port densities, demonstrating the robustness of the grouping method employed. More specifically, at a BER value of $10^{-5}$, the proposed FA-GIM scheme achieves performance gains of approximately 0.52 dB, 0.56 dB, 0.53 dB, and 0.51 dB compared to the FA-IM scheme for $W_{1}=W_{2}=$ 2.4, 4.8, 9.6, and 19.2, respectively. And the proposed FA-GIM scheme achieves BER performance gains of approximately 6.1 dB, 6.2 dB, 6.3 dB, and 6.2 dB compared to the FA-PS scheme for $W_{1}=W_{2}=$ 2.4, 4.8, 9.6, and 19.2, respectively. It is worth noting that, under the system configuration in \figref{fig:BER_W}, the proposed FA-GIM scheme achieves superior BER performance at a higher transmission rate ($B= 16$) compared to the FA-IM scheme ($B^{\prime}= 14$). This indicates that the proposed FA-GIM scheme can provide a significantly greater performance gain at the same transmission rate.

\section{Conclusion}  \label{Sec-Conc}

This paper proposes an innovative FA grouping IM scheme, namely FA-GIM, for FA-assisted MIMO systems. In existing MIMO systems, the performance of FA-IM scheme degrades due to high correlation between ports on the FA. The proposed FA-GIM scheme can mitigate the impact of spatial correlation on the system. Based on the spatial correlation model and distribution of the ports, the block grouping method is adopted. Then, the mapping functions from positions to labels of ports are established according to the grouping result. Subsequently, the theoretical ABEP upper bound for the proposed FA-GIM scheme is derived. The theoretical upper bound approaches the simulation results as the SNR increases, suggesting that the upper bound can serve as an effective theoretical tool to evaluate the system performance. Simulation results also indicate that, compared to existing schemes, the proposed FA-GIM scheme provides stable and higher performance gains in the spatially correlated channel. Specifically, the FA-GIM scheme offers superior BER performance gains as the number of receive antennas decreases, compared to the FA-IM scheme under the same transmission rate. The results also demonstrate that the FA-GIM scheme consistently provides significant BER performance gains under nearly the same transmission rates, regardless of the density of the ports, i.e., the strength of spatial correlation. This demonstrates the robustness of the proposed grouping scheme.




\bibliographystyle{IEEEtran}
\bibliography{IEEEabrv,mybib}

\end{document}